\documentclass[12pt]{article}
\usepackage{amsmath}
\usepackage{amssymb}
\usepackage{epsfig}
\usepackage{euscript}
\usepackage{fancybox}
\usepackage{color}

\def\mathswitchr#1{\relax\ifmmode{\mathrm{#1}}\else$\mathrm{#1}$\fi}

\newcommand {\pslash}{\hbox{$\not\hbox{\kern-2.3pt $p$}$}}

\newcommand {\umf}{\mathfrak{u}}

\usepackage{cite}

\usepackage{epic}

\let\sstl=\scriptscriptstyle
\def\alf1{ {\alpha\over\pi} }

\begin{document}
\begin{titlepage}
\begin{flushright}
{\bf MPI-PhT-2002-07}\\
 {\bf UTHEP-02-0301 }\\
{\bf Mar., 2002}\\
\end{flushright}
 
\begin{center}
{\Large YFS MC Approach to QCD Soft Gluon Exponentiation$^{\dagger}$
}
\end{center}

\vspace{2mm}
\begin{center}
  {\bf   B.F.L. Ward$^{a,b}$ and S. Jadach$^{c,d}$}

\vspace{2mm}
{\em $^a$Werner-Heisenberg-Institut, Max-Planck-Institut fuer Physik,
Muenchen, Germany,}\\
{\em $^b$Department of Physics and Astronomy,\\
  The University of Tennessee, Knoxville, Tennessee 37996-1200, USA,}\\
{\em $^c$CERN, Theory Division, CH-1211 Geneva 23, Switzerland,}\\
{\em $^d$Institute of Nuclear Physics,
        ul. Kawiory 26a, Krak\'ow, Poland.}
\end{center}


\vspace{5mm}
\begin{center}
{\bf   Abstract}
\end{center}
We present two things in this discussion. First, we develop and prove the
theory of the 
extension of the YFS Monte Carlo approach to higher order
$SU_{2L}\times U_1$ radiative corrections to the analogous higher
order QCD radiative corrections. Contact is made with other pioneering
soft gluon resummation theories in the literature. 
Second, semi-analytical results 
and preliminary explicit Monte Carlo data are
presented for the specific example of the processes
$p\bar{p}\rightarrow t\bar{t}+n(G)+X$ at FNAL energies, where G is a soft
gluon and the respective event generator, ttp1.0, features
realistic, event-by-event simulation of multiple, soft, finite $p_T$
gluon effects in which the infrared singularities are canceled
to all orders in $\alpha_s$. 
We comment briefly on the implications of our results
on the CDF/D0 observations and on their possible applications
to RHIC physics and to LHC physics.
\vspace{10mm}
\begin{center}
{\it Presented at the 2002 Cracow Epiphany Conference by B.F.L. Ward}
\end{center}
\vspace{10mm}
\renewcommand{\baselinestretch}{0.1}
\footnoterule
\noindent
{\footnotesize
\begin{itemize}
\item[${\dagger}$]
Work partly supported 
by the US Department of Energy Contract  DE-FG05-91ER40627
, by NATO Grant PST.CLG.977751, and by
Polish Government grant 5P03B09320.
\end{itemize}
}

\end{titlepage}

\def\Kmax{K_{\rm max}}\def\ieps{{i\epsilon}}\def\rQCD{{\rm QCD}}
\renewcommand{\theequation}{\arabic{equation}}
\font\fortssbx=cmssbx10 scaled \magstep2
\renewcommand\thepage{}
\parskip.1truein\parindent=20pt\pagenumbering{arabic}\par
\section{\bf Introduction}\label{intro}\par
The problem of soft gluon resummation is well known~\cite{sterm,cat}
and some of its many phenomenological applications are also:
\begin{itemize}
\item FNAL t\=t production cross section higher order corrections
( the current situation~\cite{cdf1,d01} has 
the experimental cross section $6.2{+1.2\atop -1.1}$
to be compared with a theoretical prediction~\cite{wilb} of $5.1\pm 0.5$)
and the attendant soft gluon uncertainty in the extracted value
of $m_t = 0.1743\pm 0.0051$TeV, where~\cite{granis} $\sim 2-3$GeV
of the the latter error could be due to soft gluon uncertainties.
\item RHIC hard scattering polarized pp scattering processes.
\item $n(G)$ production in the hard nucleon-nucleon scattering processes 
which participate in the nucleus-nucleus collisions at RHIC, where
G denotes a soft QCD gluon.
\item  $b\bar b$ and $J/\Psi$ production by hard processes at FNAL.
\item heavy vector boson production at FNAL and RHIC, etc.
\end{itemize}
For the LHC/TESLA/LC, the requirements on the corresponding theoretical
precisions will be even more demanding 
and the QCD soft $n(G)$ MC exponentiation which we discuss in the following
will be an important part of the necessary theory  --
YFS~\cite{yfs} exponentiated ${\cal O}(\alpha_s^2)L$
corrections realized on an event-by-event basis.

The results which we present also will ultimately allow us to investigate
from a different perspective some of the outstanding theoretical
issues in perturbative QCD:
\begin{itemize}
\item Treatment of the 2i(f)+n(G) phase space, where
we have in mind our usual exact treatment of this phase space
to be compared with the somewhat different approach of
Catani and Seymour~\cite{catsey}, for example -- here, f denotes
a generic fermion.
\item Approaches to resummation itself~\cite{sterm,cat1,berger,laen}.
\item No-go theorems, such as the result of Ref.~\cite{chris}
, which requires that for calculations
beyond ${\cal O}(\alpha_s)$, all initial state parton masses must vanish. 
\end{itemize}

Clearly, the results which we present herein have indeed a large 
arena for further development and application. For definiteness,
we will use the process in Fig.~1 , $\bar Q(p_1) Q(q_1)\rightarrow \bar t(p_2) t(p_1) +
G_1(k_1)\cdots G_n(k_n)$, as proto-typical process, 
\begin{figure}
\begin{center}
\setlength{\unitlength}{1mm}
\begin{picture}(160,80)
\put(-2.4, -10){\makebox(0,0)[lb]{
\epsfig{file=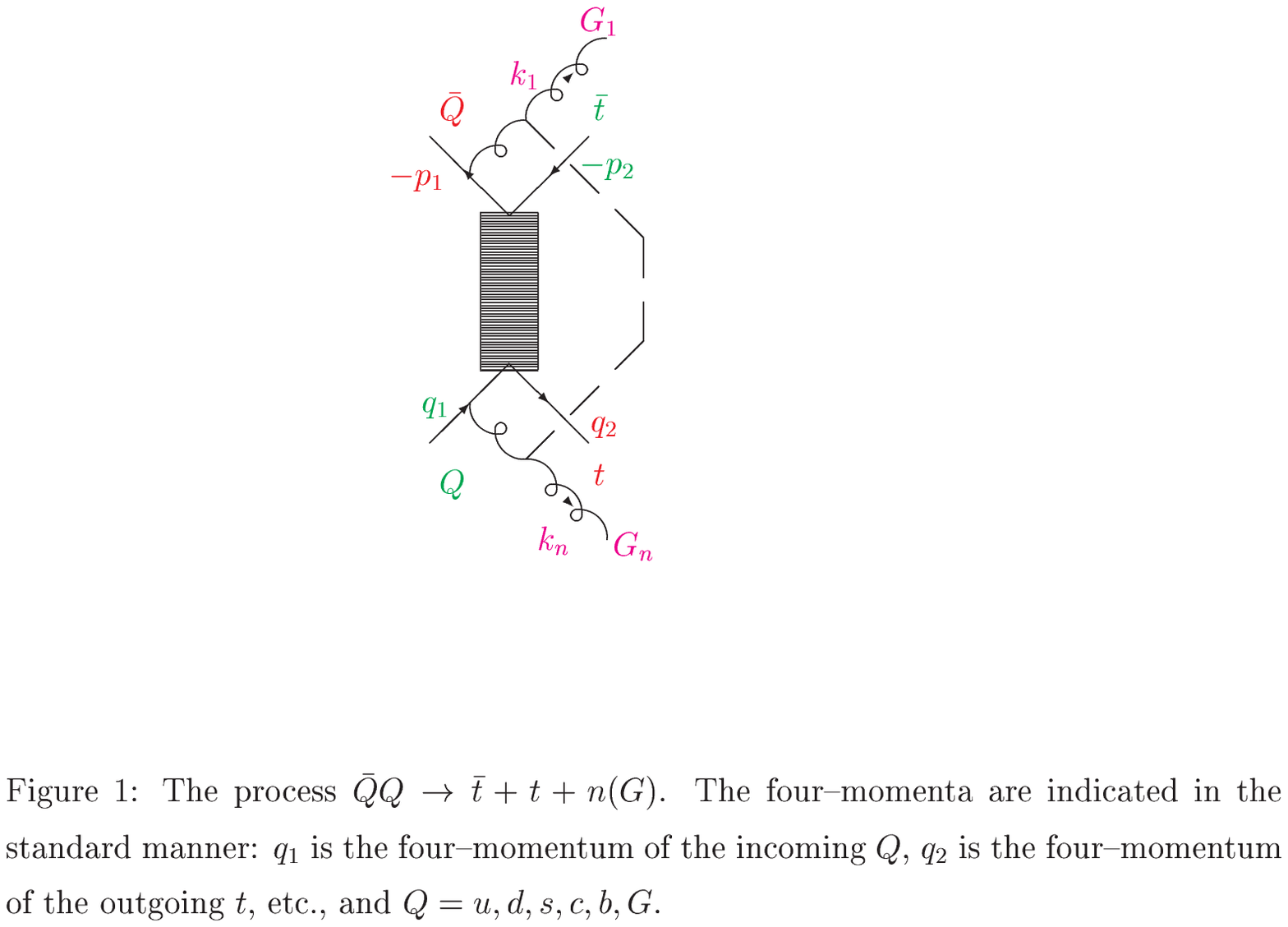,width=140mm}
}}
\end{picture}
\label{figproto.1}
\end{center} 
\end{figure}
\noindent
where we have written the
kinematics as it is illustrated in the figure. This process,
which dominates t\=t production at FNAL, contains all of the 
theoretical issues that we must face at the parton level to establish
YFS QCD soft exponentiation by MC methods -- extension to other related
processes will be immediate. For reference, let us also 
note that, in what follows, we use the GPS conventions of 
Jadach {\it et al.}~\cite{gps} for spinors and the attendant
photon and gluon polarization vectors that follow therefrom:
\begin{equation}
\label{gmapol}
  (\epsilon^\mu_\sigma(\beta))^*
     ={\bar{u}_\sigma(k) \gamma^\mu u_\sigma(\beta)
       \over \sqrt{2}\; \bar{u}_{-\sigma}(k) u_\sigma(\beta)},\quad
  (\epsilon^\mu_\sigma(\zeta))^*
     ={\bar{u}_\sigma(k) \gamma^\mu \umf_\sigma(\zeta)
       \over \sqrt{2}\; \bar{u}_{-\sigma}(k) \umf_\sigma(\zeta)},
\end{equation}
so that all phase information is strictly known in our amplitudes.
This means that, although we shall use the older EEX realization of YFS
MC exponentiation as defined in Ref.~\cite{ceex:2001}, the realization
of our results via the the newer CEEX realization of YFS exponentiation
in Ref.~\cite{ceex:2001} is also possible and is in progress~\cite{elsewh}.

We organize our presentation
as follows. In the next section, we review the application of YFS MC methods
in the EW-QED case. In Sect. 3, we prove that we can extend this 
application to the QCD theory. In Sect. 4 we illustrate
the QCD extension by applying it to t\=t production at FNAL.
Sect. 5  contains our summary remarks.
\par

\section{\bf Review of YFS Theory: An Abelian Gauge Theory Example}
In this section, for pedagogical reasons, we review the 
application of YFS exponentiation to the prototypical
Abelian gauge theory example of $e^+e^-\rightarrow \bar f f + n(\gamma)$
so that we set the stage for our proof of the extension of this
theory to the prototypical non-Abelian gauge theory example of
$q\bar q\rightarrow t\bar t+ n(G)$ in QCD in the next section.
Specifically, in Refs.~\cite{bflw,sjw1990,sjw1992} we have shown that
the process $e^+(p_1)e^-(q_1)\rightarrow \bar{f}(p_2) f(q_2) +n(\gamma)(k_1,\cdots,k_n)$,
in the 
renormalization group improved YFS theory~\cite{bflw}, is represented by
\begin{equation}
d\sigma_{exp}=e^{2\alpha\,Re\,B+2\alpha\,
\tilde B}\sum_{n=0}^\infty{1\over n!}\int\prod_{j=1}^n{d^3k_j\over k_j^0
}\int {d^4y\over(2\pi)^4}e^{iy(p_1+q_1-p_2-q_2-\sum_jk_j)+D}\nonumber\cr
\qquad\bar\beta_n(k_1,\dots,k_n){d^3p_2d^3q_2\over p_2^0q_2^0}
\label{eqone}\end{equation}
where the
real infrared function $\tilde B$ and the virtual infrared function $B$ are
given in Refs.~\cite{yfs,sjw1990,BHLUMI-89,BHLUMI-92,ward,yfsww1}
, and where we note the usual connections
\[2\alpha\,\tilde B = \int^{k\le K_{max}}{d^3k\over k_0}\tilde S(k)\nonumber\]
\begin{equation}D=\int d^3k{\tilde S(k)\over k^0}\left(e^{-iy\cdot k}-\theta(K_{
max}-k)\right)\label{eqtwo}\end{equation} for the standard YFS infrared emission
factor\begin{equation}\tilde S(k)= {\alpha\over4\pi^2}\left[Q_fQ_{
{\llap{\phantom f}^{\sstl(}\bar f^{\sstl^)}{}}'}\left({p_1\over p_1\cdot k}-{q_1
\over q_1\cdot k}\right)^2+(\dots)\right]\label{eqthree}\end{equation} if $Q_f$
is the electric charge of $f$ in units of the positron charge. 
For example, the YFS hard photon residuals $\bar\beta_i$ in (\ref{eqone}), 
$i=0,1,2$,
are given in Ref.~\cite{BHLUMI-96} for BHLUMI 4.04 so that this latter
event generator calculates the YFS exponentiated exact ${\cal O}(\alpha)$
and LL ${\cal O}(\alpha^2)$
cross section for Bhabha scattering using a
corresponding Monte Carlo realization
of (\ref{eqone}). In the remainder of this paper, we will develop
and apply the analogous theoretical paradigm to the 
prototypical QCD higher order radiative corrections problem
for $q\bar q\rightarrow t\bar t +n(G)$. 
\par

\section{Extension to non-Abelian Gauge Theories: Proof}

Specifically, in Refs.~\cite{delaney},
we have analyzed how in the special case of Born level color exchange
one applies the YFS theory to QCD by extending
the respective YFS
IR singularity analysis to QCD to all orders in $\alpha_s$.
Here, we will take a somewhat different approach.
We want to focus on the YFS theory as a general re-arrangement of
renormalized perturbation theory based on its IR
behavior, just as the renormalization group
is a general property of renormalized perturbation theory based on its
UV(ultra-violet) behavior.
Thus, here we will keep our arguments
entirely general from the outset, so that it will be immediate
that our result applies to any renormalized perturbation theory
in which the cross section under study is finite.\par

Let the amplitude for the emission of $n$ real gluons in our proto-typical
subprocess, 
$q_i^\alpha + \bar q_i^{\bar\alpha} \rightarrow t^\gamma
\bar t^{\bar\gamma} + n(G)$, where $\alpha, {\bar\alpha}, \gamma$, and
${\bar\gamma}$ are color indices, be
represented by
\begin{equation}
{\cal M}^{(n)\alpha{\bar\alpha}}_{\gamma{\bar\gamma}} 
 = \sum_{\ell}M^{(n)\alpha{\bar\alpha}}_{\gamma{\bar\gamma}\ell},
\label{subp1}
\end{equation}
$M^{(n)}_{\ell}$ is the contribution to 
${\cal M}^{(n)}$
from Feynman diagrams with ${\ell}$ virtual loops.
Symmetrization yields
\begin{eqnarray}
   M^{(n)}_\ell = {1\over {\ell!}}\int\prod_{j=1}^{\ell}{d^4k_j\over{(2\pi)^4(
 k_j^2-\lambda^2+i\epsilon)}}\rho^{(n)}_\ell(k_1,\cdots,k_\ell),
\label{subp2}
\end{eqnarray}
where this last equation defines $\rho^{(n)}_\ell$ as a symmetric
function of its arguments arguments $k_1,...,k_{\ell}$.
$\lambda$ will be our infrared gluon regulator mass for IR singularities;
n-dimensional regularization of the 't Hooft-Veltman~\cite{tHvelt} type is also
possible as we shall see.

We now define the virtual IR emission factor $S_{QCD}(k)$ for a gluon
of 4-momentum $k$, for the $k\rightarrow 0$ regime of the
respective 4-dimensional loop integration as in (\ref{subp2}), such that
\begin{equation}
\lim_{k\rightarrow 0}k^2\left( \rho^{(n)\alpha{\bar\alpha}}_{\gamma{\bar\gamma}1}(k)
-S_{QCD}(k)\rho^{(n)\alpha{\bar\alpha}}_{\gamma{\bar\gamma}0}\right)|_{\alpha\neq\bar\alpha\neq\gamma\neq\alpha} = 0, 
\end{equation}
where we have now introduced the Born level color exchange condition as
$\alpha\neq\bar\alpha\neq\gamma\neq\alpha$ for definiteness.
(Henceforth, when we refer to $k\rightarrow 0$ gluons we are always referring
for virtual gluons to the corresponding regime of the 4-dimensional
loop integration in the computation of $ M^{(n)}_\ell$.)

In Ref.~\cite{delaney},
we have calculated $S_{QCD}(k)$
using the running quark masses to regulate its collinear
mass singularities, for example;  n-dimensional regularization of
the 't Hooft-Veltman type is also possible for
these mass singularities and we will also illustrate this presently.

We stress that $S_{QCD}(k)$
has a freedom in it corresponding to the fact that any
function $\Delta S_{QCD}(k)$ which has the property that
$\lim_{k\rightarrow 0} k^2\Delta S_{QCD}(k)\rho^{(n)}_0=0$
may be added to it. 

Since the virtual gluons in $\rho^{(n)}_{\ell}$
are all on equal footing by the symmetry of this function,
if we look at gluon $\ell$, for example, we may write
, for $k_{\ell} \rightarrow (0,0,0,0)\equiv O$ while the
remaining $k_i$ are fixed away from $O$, the representation
\begin{equation}
 \rho^{(n)}_\ell = S_{QCD}(k_\ell)*\rho^{(n)}_{\ell-1}(k_1,\cdots,k_{\ell-1})
 +\beta^1_\ell(k_1,\cdots,k_{\ell-1};k_\ell)
\label{subp3}
\end{equation}
where the residual amplitude $\beta^1_\ell(k_1,\cdots,k_{\ell-1};k_\ell)$
will now be taken as defined by this last equation. It has two nice 
properties:
\begin{itemize} 
\item it is symmetric in its first $\ell - 1$ 
arguments 
\item the
IR singularities for gluon $\ell$ {\it that are contained in 
$S_{QCD}(k_\ell)$} are no longer contained in it. 
\end{itemize}

We do not at this point
discuss whether there are any further remaining IR singularities
for gluon $\ell$ in $\beta^1_\ell(k_1,\cdots,k_{\ell-1};k_\ell)$.
In an Abelian gauge theory like QED, as has been shown by Yennie, Frautschi
and Suura in Ref.~\cite{yfs},
there would not be any further
such singularities; for a non-Abelian gauge theory like QCD, this point
requires further discussion and 
we will come back to this point presently.

We rather now stress that
if we apply the representation (\ref{subp3}) again we may write
{\small
\begin{eqnarray}
   \rho^{(n)}_\ell = S_{QCD}(k_\ell)S_{QCD}(k_{\ell-1})* \rho^{(n)}_{\ell-2}(k_1,\cdots,
 k_{\ell-2})\nonumber \\
 + S_{QCD}(k_\ell)\beta^1_{\ell-1}(k_1,\cdots,k_{\ell-2};k_{\ell-1})
\nonumber \\
+ S_{QCD}(k_{\ell-1})\beta^1_{\ell-1}(k_1,\cdots,k_{\ell-2};k_\ell)\nonumber \\
+ \beta^2_\ell(k_1,\cdots,k_{\ell-2};k_{\ell-1},k_\ell),
\label{subp4}
\end{eqnarray} 
where this last equation serves to define the function
$\beta^2_\ell(k_1,\cdots,k_{\ell-2};k_{\ell-1},k_\ell)$. It has two nice
properties:
\begin{itemize} 
\item it is symmetric in its first $\ell -2$ arguments
and in its last two arguments $k_{\ell-1},k_\ell$ 
\item the infrared
singularities for gluons $\ell-1$ and $\ell$ that are contained in
$S_{QCD}(k_{\ell-1})$ and $S_{QCD}(k_\ell)$ are no longer contained in
it. 
\end{itemize}

Continuing in this way, with repeated application of (\ref{subp3}),
we get finally the rigorous, exact rearrangement of the 
contributions to $\rho^{(n)}_\ell$ as
\begin{eqnarray}
\rho^{(n)}_\ell = S_{QCD}(k_1)\cdots S_{QCD}(k_\ell)\beta^0_0+\sum_{i=1}^\ell\prod_{j\neq i}
S_{QCD}(k_j)\beta^1_1(k_i) +\cdots \nonumber\\
+\beta^\ell_\ell(k_1,\cdots,k_\ell),
\label{subp5}
\end{eqnarray}
where the virtual gluon residuals  $\beta^i_i(k'_1,\cdots,k'_i)$
have two nice properties:
\begin{itemize} 
\item they are symmetric functions of their
arguments 
\item they do not contain any of the IR singularities which
are contained in the product 
$S_{QCD}(k'_1)\cdots S_{QCD}(k'_i)$.
\end{itemize}

Henceforth, we denote  
$\beta^i_i$ as the function $\beta_i$ for reasons of pedagogy.
We can not stress too much that (\ref{subp5}) is an {\it exact}
rearrangement of the contributions of the Feynman diagrams which
contribute to $\rho^{(n)}_\ell$; it involves no approximations.
Here also it is important to note that we do no enter into the
question of the absolute convergence of these Feynman diagrams
from the standpoint of constructive field theory. Yennie,
Frautschi and Suura~\cite{yfs} have already stressed that Feynman diagrammatic
perturbation theory is non-rigorous from this standpoint.
What we claim is that the relationship between the YFS expansion and the
usual perturbative Feynman diagrammatic expansion is itself rigorous
even though neither of the two expansions themselves is rigorous.
\par

Introducing (\ref{subp5}) into (\ref{subp1}) yields
a representation similar to that of YFS, and we
will call it a ``YFS representation'', 
\begin{equation}
{\cal M}^{(n)} = exp(\alpha_sB_{QCD})\sum_{j=0}^\infty{\sf m}^{(n)}_j,
\label{yfsrepv}
\end{equation}
where we have defined
\begin{equation}
\alpha_s(Q)B_{QCD} = \int {d^4k\over (k^2-\lambda^2+i\epsilon)}S_{QCD}(k)
\label{vbfn}
\end{equation}
and
\begin{equation}
 {\sf m}^{(n)}_j = {1\over {j!}}\int\prod_{i=1}^j{d^4k_i\over k_i^2}
       \beta_j(k_1,\cdots,k_j). 
\label{irfreev}
\end{equation}
We say that (\ref{yfsrepv}) is similar to the respective result
of Yennie, Frautschi and Suura in Ref.~\cite{yfs}
and is not identical to it because we have not
proved that the functions $\beta_i(k_1,...,k_i)$ are completely
free of virtual IR singularities. What have shown is that they do not
contain the IR singularities in the product 
$S_{QCD}(k_1)\cdots S_{QCD}(k_i)$ so that
${\sf m}^{(n)}_j$ does not contain the virtual IR divergences
generated by this product when it is integrated over the respective
4j-dimensional j-virtual gluon phase space. In an Abelian gauge theory,
there are no other possible virtual IR divergences; in the non-Abelian
gauge theory that we treat here, such additional IR divergences
are possible; but, the 
result (\ref{yfsrepv}) does have an improved
IR divergence structure over (\ref{subp1}) in that all of the
IR singularities associated with $S_{QCD}(k)$ are explicitly
removed from the sum over the virtual IR improved loop contributions
${\sf m}^{(n)}_j$ to all orders in $\alpha_s(Q)$.\par

Turning now to the analogous rearrangement of the real IR singularities in
the differential cross section associated with the ${\cal M}^{(n)}$,
we first note that we may write this cross section as follows
according the standard methods
\begin{eqnarray}
  d\hat\sigma^n = {e^{2\alpha_sReB_{QCD}}\over {n !}}\int\prod_{m=1}^n
{d^3k_m\over (k_m^2+\lambda^2)^{1/2}}\delta(P_1+Q_1-P_2-Q_2-\sum_{i=1}^nk^0_i)
\nonumber\\       
\bar\rho^{(n)}(P_1,P_2,Q_1,Q_2,k_1,\cdots,k_n)
{d^3P_2d^3Q_2\over P^0_2 Q^0_2},
\label{diff1}
\end{eqnarray}
where we have defined
\begin{equation}
\bar\rho^{(n)}(P_1,P_2,Q_1,Q_2,k_1,\cdots,k_n)=
\sum_{color,spin} \|\sum_{j=0}^\infty{\sf m}^{(n)}_j\|^2
\label{diff2}
\end{equation}}
in the incoming q\=q cms system
and we have absorbed the remaining kinematical factors for
the initial state flux, spin and color averages into the
normalization of the amplitudes ${\cal M}^{(n)}$ for reasons of
pedagogy so that the $\bar\rho^{(n)}$ are averaged over initial spins
and colors and summed over final spins and colors.
We now proceed in complete analogy with the discussion
of $\rho^{(n)}_\ell$ above. \par

Specifically, 
for the functions $\bar\rho^{(n)}(p_1,p_2,q_1,q_2,k_1,\cdots,k_n)
\equiv \bar\rho^{(n)}(k_1,\cdots,k_n)$ which are symmetric functions
of their arguments $k_1,\cdots,k_n$, we define first, for $n=1$, 
\begin{equation}
\lim_{|\vec{k}|\rightarrow 0}\vec{k}^2\left(\bar\rho^{(1)}(k)
-\tilde S_{QCD}(k)\bar\rho^{(0)}\right) = 0, 
\label{realS}
\end{equation}
where the real infrared function $\tilde S_{QCD}(k)$ is rigorously
defined by this last equation and is explicitly computed in
DeLaney {\it et al.}~\cite{delaney}; 
like its virtual counterpart $S_{QCD}(k)$
it has a freedom in it in that any function $\Delta\tilde S_{QCD}(k)$
with the property that 
$\lim_{|\vec{k}|\rightarrow 0}\vec{k}^2\Delta\tilde S_{QCD}(k)=0$ 
may be added to it without affecting the defining relation (\ref{realS}).
 
As we show in Ref.~\cite{elsewh}, we can again repeat the 
analogous arguments of Ref.~\cite{yfs} to get 
the ``YFS-like'' result {\small 
\begin{equation}
\begin{split}
d\hat\sigma_{\rm exp}&= \sum_n d\hat\sigma^n \\
         &=e^{\rm SUM_{IR}(QCD)}\sum_{n=0}^\infty\int\prod_{j=1}^n{d^3
k_j\over k_j}\int{d^4y\over(2\pi)^4}e^{iy\cdot(P_1+P_2-Q_1-Q_2-\sum k_j)+
D_\rQCD}\\
&*\bar\beta_n(k_1,\ldots,k_n){d^3P_2\over P_2^{\,0}}{d^3Q_2\over
Q_2^{\,0}}
\end{split}
\label{subp10}
\end{equation}
with 
\[ {SUM}_{IR}(QCD)=2\alpha_s ReB_{QCD}+2\alpha_s\tilde B_{QCD}(\Kmax),\]
\[ 2\alpha_s\tilde B_{QCD}(\Kmax)=\int{d^3k\over k^o}\tilde S_\rQCD(k)
\theta(\Kmax-k),\]
 \begin{equation} D_\rQCD=\int{d^3k\over k}\tilde S_\rQCD(k)
\left[e^{-iy\cdot k}-\theta(\Kmax-k)\right],\label{subp11}\end{equation}
\[{1\over 2}\bar\beta_0=d\sigma^{\rm(1-loop)}-2\alpha_s{\rm Re}B_\rQCD d\sigma_B,\]
\begin{equation}{1\over 2}\bar\beta_1=d\sigma^{B1}-\tilde S_\rQCD(k)d\sigma_B,\quad\ldots\label{subp12}\end{equation}
where the $\bar\beta_n$ are the QCD hard gluon residuals defined above; they
are the non-Abelian analogs of the hard photon residuals defined by YFS.
Here, for illustration, we have recorded the
relationship between the $\bar\beta_n$, $n=0,1$ through ${\cal O}(\alpha_s)$
and the exact one-loop and single bremsstrahlung cross sections,
$d\sigma^{\rm(1-loop)}$, $d\sigma^{B1}$, respectively, where the latter
may be taken from Nason {\it et al.}~\cite{nason}
and
Beenakker {\it et al.}~\cite{been1}
We stress two things about the right-hand side of
(\ref{subp10}) :
\begin{itemize}
\item It does not depend on the dummy parameter $K_{max}$ which has been
introduced for cancellation of the infrared divergences in 
$SUM_{IR}(QCD)$ to all orders in $\alpha_s(Q)$ where $Q$ is the hard
scale in the parton scattering process under study here.
\item Its analog can aslo be derived in our new CEEX~\cite{ceex:2001} 
format.
\end{itemize}

We now
return to the property of (\ref{subp10}) that distinguishes it
from the Abelian result derived by YFS -- namely, the fact
that, owing to its non-Abelian gauge theory origins, it may very well be
that there are infrared divergences in the $\bar\beta_n$ which were not
removed into the $S_{QCD},\tilde S_{QCD}$ when these infrared functions
were isolated in our derivation of (\ref{subp10}).}\par

More precisely, the left-hand side of (\ref{subp10}) is the fundamental
reduced parton cross section and it should be infrared finite or else
the entire QCD parton model has to be abandoned. 

There is an observation
in the literature~\cite{chris} that 
unless we use
some collinear mass singularity regulator other than the incoming
light quark running masses, the reduced parton cross section
on the left-hand side of (\ref{subp10}) diverges in the infrared
regime at ${\cal O}(\alpha_s^2(Q))$. We do not
go into this issue here but use n-dimensional methods
to regulate such divergences while setting the quark masses
to zero as that is an excellent approximation for the
light quarks at FNAL energies -- we take this issue up elsewhere.\par

From the infrared finiteness of the left-hand side of (\ref{subp10})
and the infrared finiteness of $SUM_{IR}(QCD)$, it follows
that the quantity 
\[d\bar{\hat\sigma}_{\rm exp}\equiv 
\exp[{\rm SUM_{IR}(QCD)}]d\hat\sigma_{\rm exp}\]
must also be infrared finite to all orders in $\alpha_s$.

As we assume the QCD theory makes sense in some neighborhood of the
origin for $\alpha_s$, we conclude that each order in $\alpha_s$
must make an infrared finite contribution to $d\bar{\hat\sigma}_{\rm exp}$. 
At ${\cal O}(\alpha_s^0(Q))$ , the only contribution to 
$d\bar{\hat\sigma}_{\rm exp}$ is the respective Born cross section
given by $\bar\beta^{(0)}_0$ in (\ref{subp10}) and it
is obviously infrared finite, where we use henceforth the notation
$\bar\beta^{(\ell)}_n$ to denote the ${\cal O}(\alpha_s^{\ell}(Q))$
part of $\bar\beta_n$. Thus, we conclude that
the lowest hard gluon residual $\bar\beta^{(0)}_0$ is infrared finite.

Let us now define the left-over 
non-Abelian infrared divergence part
of each contribution $\bar\beta^{(\ell)}_n$ via
\[ \bar\beta^{(\ell)}_n= \tilde{\bar\beta}^{(\ell)}_n + D\bar\beta^{(\ell)}_n\]
where the new function $\tilde{\bar\beta}^{(\ell)}_n$ is now completely
free of any infrared divergences and the function $D\bar\beta^{(\ell)}_n$
contains all left-over infrared divergences in $\bar\beta^{(\ell)}_n$
which are of non-Abelian origin and is normalized to vanish in the
Abelian limit $f_{abc}\rightarrow 0$ where $f_{abc}$ are the group
structure constants.

Further, we define $D\bar\beta^{(\ell)}_n$
by a minimal subtraction of the respective IR divergences in it
so that it only contains the actual pole and transcendental
constants, $1/\epsilon -C_E$ for $\epsilon=2-d/2$, where $d$ is
the dimension of space-time, in dimensional regularization
or $\ln \lambda^2$ in the gluon mass regularization. Here, $C_E$ is
Euler's constant. 

For definiteness, we 
write this out explicitly so that there can be no confusion about what
we mean
\[\int dPh\; D\bar\beta^{(\ell)}_n\equiv \sum_{i=1}^{n+\ell} d^{n,\ell}_i\ln^i(\lambda^2)\] 
where the coefficient functions $d^{n,\ell}_i$ are independent of $\lambda$ for
$\lambda\rightarrow 0$ and $dPh$ is the respective n-gluon Lorentz
invariant phase space.

At ${\cal O}(\alpha_s^n(Q))$, the IR finiteness
of the contribution to $d\bar{\hat\sigma}_{\rm exp}$ then requires
the contribution {\small
\begin{eqnarray}
d\bar{\hat\sigma}^{(n)}_{\rm exp} \equiv
\int\sum_{\ell=0}^n\frac{1}{\ell!}\prod_{j=1}^{\ell}\int_{k^0_j\ge K_{max}}{d^3
k_j\over k_j}\tilde S_{QCD}(k_j)\sum_{i=0}^{n-\ell}\frac{1}{i!}\prod_{j=\ell+1}^{\ell+i}
\nonumber\\
\int{d^3k_j\over k_j}
\bar\beta^{(n-\ell-i)}_i(k_{\ell+1},\ldots,k_{\ell+i}){d^3P_2\over P_2^{\,0}}{d^3Q_2\over
Q_2^{\,0}}\label{subp13}
\end{eqnarray}}
to be finite. 

From this it follows that {\small
\begin{eqnarray}
Dd\bar{\hat\sigma}^{(n)}_{\rm exp} \equiv
\int\sum_{\ell=0}^n\frac{1}{\ell!}\prod_{j=1}^{\ell}\int_{k^0_j\ge K_{max}}{d^3
k_j\over k_j}\tilde S_{QCD}(k_j)\sum_{i=0}^{n-\ell}\frac{1}{i!}\prod_{j=\ell+1}^{\ell+i}
\nonumber\\
\int{d^3k_j\over k_j} D\bar{\beta}^{(n-\ell-i)}_i(k_{\ell+1},\ldots,k_{\ell+i}){d^3P_2\over P_2^{\,0}}{d^3Q_2\over
Q_2^{\,0}}\label{subp14}
\end{eqnarray}}
is finite. Since the integration region for the final particles is
arbitrary, the independent powers of the IR regulator $\ln(\lambda^2)$ in this
last equation must give vanishing contributions.
This means that
we can drop the $D\bar\beta^{(\ell)}_n$ from our result
(\ref{subp10}) because they do not make a net contribution to the final
parton cross section $\hat\sigma_{\rm exp}$. We thus finally arrive
at the new rigorous result{\small
\begin{equation}
\begin{split}
d\hat\sigma_{\rm exp}&= \sum_n d\hat\sigma^n \\
         &=e^{\rm SUM_{IR}(QCD)}\sum_{n=0}^\infty\int\prod_{j=1}^n{d^3
k_j\over k_j}\int{d^4y\over(2\pi)^4}e^{iy\cdot(P_1+P_2-Q_1-Q_2-\sum k_j)+
D_\rQCD}\\
&*\tilde{\bar\beta}_n(k_1,\ldots,k_n){d^3P_2\over P_2^{\,0}}{d^3Q_2\over
Q_2^{\,0}}
\end{split}
\label{subp15}
\end{equation}
where now the hard gluon residuals 
$\tilde{\bar\beta}_n(k_1,\ldots,k_n)$
defined by 
\[\tilde{\bar\beta}_n(k_1,\ldots,k_n= \sum_{\ell=0}^\infty 
\tilde{\bar\beta}^{(\ell)}_n(k_1,\ldots,k_n)\]
are free of all infrared divergences to all 
orders in $\alpha_s(Q)$.
This is a basic result of this paper. 
We note here that, contrary to what we claimed in the Appendix of the
first paper in Refs.~\cite{delaney}, the arguments in Refs.~\cite{delaney}
are not sufficient to derive the respective analog of eq.(\ref{subp15});
for, they did not really expose the compensation between
the left over genuine nonAbelian IR virtual and real singularities
between $\int dPh\bar\beta_n$ and $\int dPh\bar\beta_{n+1}$ respectively
that really distinguishes
QCD from QED, where no such compensation occurs in the $\bar\beta_n$
residuals for QED. Further discussion of the many implications of
eq.(\ref{subp15}) is given elsewhere~\cite{elsewh}.
\par

\section{YFS Exponentiated QCD Corrections to t\= t Production at High Energies}

We shall realize the result above as it is
applied to the process in Fig. 1 at high energies by the Monte Carlo
event generator methods of Ref.~\cite{sjw1990,sjw1992}. The first
application of these methods to QCD processes has already been
reported in DeLaney {\it et al.}~\cite{delaney}.
Here, we shall apply the general
result in the latter reference to the t\= t production at the
parton level and show how to synthesize that parton level result
with the realistic parton distributions to arrive at an event
generator for $p\bar p\rightarrow t\bar t+X$. Similar results
will hold for pp incoming states.
Sample MC data will
be illustrated. We refer to the respective MC event generator as 
ttp1.0. It is in the EEX format but a CEEX version
is imminent. It will be available from the authors soon.
\par

The starting point in our MC realization of basic result
for t\= t production at the hadron level is its realization
for the respective parton level processes. In Ref.~\cite{delaney}
we have shown how to construct such parton level event generators
for the processes such as $q\bar q\rightarrow t\bar t+n(G)$
where $q$ is a light quark. Here, we extend these results
to the process $G+G\rightarrow t\bar t+n(G)$. This is
presented in detail elsewhere~\cite{elsewh}.
\par

Following the procedures in Ref.~\cite{delaney}, 
we then use the Monte Carlo algorithm presented
in S. Jadach {\it et al.}~\cite{sjw1990,BHLUMI-89} to realize the 
result derived in the previous section
for both the  $q\bar q\rightarrow t\bar t+n(G)$ 
and $GG\rightarrow t\bar t+n(G)$ subprocesses
on an event-by-event basis in which infrared singularities are now canceled
to all orders in $\alpha_s$.\par

In order to apply these parton level results to the desired hadron
level cross section $\sigma(p\bar p\rightarrow t\bar t+X)$, 
we use the standard formula
\begin{equation}
\sigma(p\bar p\rightarrow t\bar t+X)=\int\sum_{i,j}F_i(x_i)\bar F_j(x_j)
d\hat\sigma'_{{\rm exp},ij}dx_idx_j
\label{eqsigtot}
\end{equation}
where $F_i(\bar F_j)$ is the structure function of parton $i(j)$ in p(\=p)
and where $\hat\sigma'_{{\rm exp},ij}$ is the result derived above
for the t\=t production subprocess with the incoming parton-i,parton-j
initial state 
when the DGLAP synthesization procedure presented in Ref.~\cite{dglapsyn}
is applied to it to avoid over-counting
resummation effects already included in the structure function
DGLAP evolution. 
In operational terms, the DGLAP synthesized result
$\hat\sigma'_{{\rm exp},ij}$ is obtained from the result
$\hat\sigma_{{\rm exp},ij}$ of the previous section
by substituting the functions $\tilde S^{nls}_{QCD},B^{nls}_{QCD}$ for $\tilde S_{QCD}, B_{QCD}$, respectively, where the former functions
are given in Ref.~\cite{dglapsyn} and do not contain big logs already
contained in the structure functions, for example.\par

The formula in (\ref{eqsigtot})
we have realized by MC methods by extending the MC realizations
of (\ref{subp15}) for the $q\bar q$ and $GG$ subprocesses
to the respective realizations of $d\hat\sigma'_{{\rm exp},ij},~ij=q\bar q,~GG$
in the presence of the additional two-dimensional structure function
distribution in a standard way which is already illustrated for example
in Ref.~\cite{delaney2}.
In this way, we arrive at the MC event generator
ttp1.0, which features YFS-style exponentiated
multiple soft gluon radiative effects in p\=p$\rightarrow$t\=t$+X$
on an event-by-event basis
in which infrared singularities are canceled to all orders in
$\alpha_s$. We now illustrate its application
with both semi-analytical results and with
with some preliminary sample MC data at FNAL energies.\par

As a first check of the approach, we have realized the formula
in (\ref{subp15}) by semi-analytical methods. This is useful for a number of
reasons. For example, it allows us to check the normalization of our
work with that of other resummed calculations in the
literature. An important outcome is an estimate of the size
of the ${\cal O}(\alpha_s^n),n\ge 2$ contribution to $\sigma(t\bar t)$
at FNAL~\cite{wilb}.

Specifically, as we show in Ref.~\cite{mpa-ttbar}, our
semi-analytical result for the ratio of the YFS exponentiated
cross section to the corresponding Born cross section is given by
\begin{eqnarray}
r^{nls}_{exp} &=& 
    \exp\left\{ {\alpha_s\over\pi}[(2C_F-{1\over 2}C_A)
         {\pi^2\over 3}-{1\over 2}C_F] \right\}
\nonumber \\
      &=&  
{\begin{cases}
       1.086,& \text{$\alpha_s=\alpha_s(\sqrt s)$},\\
       1.103,& \text{$\alpha_s=\alpha_s(2m_t)$},\\
       1.110,& \text{$\alpha_s=\alpha_s(m_t)$}.\\
  \end{cases}} 
\label{rexphl}
\end{eqnarray}
This implies that the prediction for the contribution to
$\sigma(p\bar p\rightarrow t\bar t)$ from $({\cal O}(\alpha_s^n), n\ge 2)$ is
\begin{eqnarray}
\delta\sigma(p\bar p\rightarrow t\bar t)^{exp} = \int\sum_{i,j}D_i(x_i)\bar D_j(x_j)
(r^{nls}_{exp}-1-{\alpha_s\over \pi}[(2C_{ij}-{1\over 2}C_A)
         {\pi^2\over 3}
\nonumber\\
-{1\over 2}C_{ij}])d\hat{\sigma}_B(x_ix_js)
dx_idx_j ,
\label{eqa10}
\end{eqnarray}}
where
\begin{eqnarray}
C_{ij} =
\begin{cases}
         C_F, &\text{$ij=q\bar q,~\bar q q$},\\  
         C_A, &\text{$ij=GG$}.\\
\end{cases}
\label{eqa11}
\end{eqnarray}
From this we get that  $({\cal O}(\alpha_s^n), n\ge 2)$ contributes
.006-.008 of the NLO cross section, in agreement with 
Catani {\it et al.} in Ref.~\cite{cat1}.\par

We have also generated preliminary MC data using the 
proto-typical version of ttp1.0. A detail presentation of such data
will appear elsewhere~\cite{elsewh}. Here, in view of its importance,
we show preliminary data for the $p_T$ spectrum from ttp1.0.
This is shown in Fig. 2.
\newpage
\begin{center}
\setlength{\unitlength}{1mm}
\begin{picture}(100,60)
\put(15, -45){\makebox(0,0)[lb]{
\epsfig{file=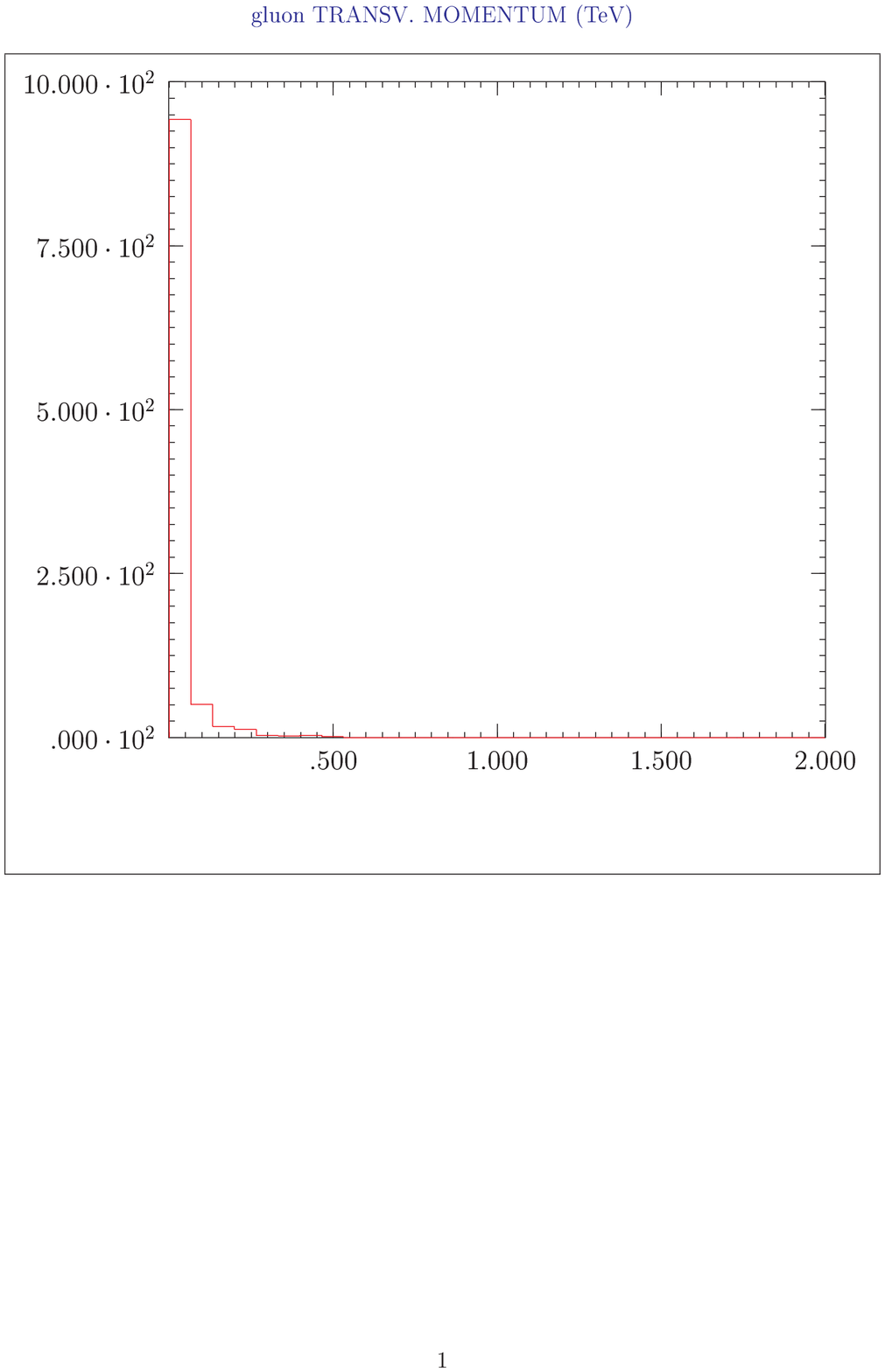,width=75mm}
}}
\end{picture}
\end{center}
Figure 2. $n(G)$ transverse momentum distribution for in p\=p$\rightarrow$
t\=t +X+$n(G)$. The result is a preliminary.

\noindent
What we see is that, indeed, the soft $n(G)$ effects do modulate significantly
the $p_T$ spectrum on the scale of a few GeV -- they must be taken into
account properly to understand observables derived from them on this level.
\par

\section{Conclusions}

We conclude that YFS theory, both the EEX and CEEX formulations,
extends to QCD. In our work, a full MC event generator realization of this
extension is near. Our early semi-analytical results
agree with the literature on t\=t production at FNAL energies.
Preliminary MC data show that the $n(G)$ $p_T$ is significant
in the latter process. We are currently pursuing the attendant
ramifications for LHC and RHIC as well.\par

\section*{Acknowledgements}\label{acks}
One of us (B.F.L.W.) thanks Prof. S. Bethke for the support and kind
hospitality of the MPI, Munich, during the final stages of this work.
The authors also thank Profs.~G.~Altarelli and Prof. Wolf-Dieter Schlatter
for the support and kind hospitality 
of CERN while a part of this work was completed.
These same two authors also thank Profs.~F.~Gilman and W.~Bardeen of the
former SSCL for their kind hospitality while this work was in its development
stages.\par

\newpage
\section*{Figure Captions}
\noindent
Figure 1. The process $\bar{Q}
    Q      \rightarrow        \bar{t}
                 +      t    + n(G)$.
The four--momenta are indicated in the standard manner: $q_1$ 
is the
four--momentum of the incoming $Q$, $q_2$ 
is the four--momentum of the outgoing $t$, etc.,
and $Q = u,d,s,c,b,G.$\par
\noindent
Figure 2. $n(G)$ transverse momentum distribution for in p\=p$\rightarrow$
t\=t +X+$n(G)$. The result is a preliminary. 

\begin{thebibliography}{99}
\bibitem{sterm}
G. Sterman, Nucl. Phys. {\bf B281} (1987) 310.
\bibitem{cat}
S. Catani and L. Trentadue, 
Nucl. Phys. {\bf B327} (1989) 323; {\em ibid.}~{\bf B353} (1991) 183.
\bibitem{cdf1}
S. R. Blusk, in {\em Proc. ICHEP2000}, ed. C.S. Lim and T. Yamanaka,
 ( World Scientific, Singapore, 2001) p. 811.
\bibitem{d01}
D. Chakraborty, in {\em Proc. ICHEP2000}, ed. C.S. Lim and T. Yamanaka,
 ( World Scientific, Singapore, 2001) p. 814.
\bibitem{wilb} S. R. Willenbrock, in {\em Proc. RADCOR2000, Carmel, 2000}, eds.
H. Haber and S. Brodsky, eConf C000911, (2001) 379;
Rev. Mod. Phys. {\bf 72} (2000) 1141.
\bibitem{granis}P. Granis, private communication.
\bibitem{yfs}
D.~R.~Yennie, S.~C.~Frautschi, and H.~Suura, Ann. 
Phys. {\bf 13} (1961) 379;\newline
see also K.~T.~Mahanthappa, Phys.~Rev.~{\bf 126} (1962) 329, for a related analysis.
\bibitem{catsey} S. Catani and M.H. Seymour,
Nucl. Phys. {\bf B485} (1997) 291; {\bf ibid.}{\bf B510} (1997) 503;
Phys. Lett. {\bf B378} (1996) 287.
\bibitem{cat1}S. Catani {\it et al.}, Phys. Lett. {\bf B378} (1996) 329;
Nucl. Phys. {\bf B478} (1996) 273.
\bibitem{berger} 
E. Berger and H. Contopanagos, Phys. Rev. D{\bf 54} (1996) 3085.
\bibitem{laen}E.~Laenen, J.~Smith, and W.~van Neerven,
Phys. Lett. {\bf B321} (1994) 254.
\bibitem{chris}C. Di'Lieto, S. Gendron, I.G. Halliday and Christopher T. Sachrajda, Nucl. Phys. {\bf B183} (1981) 223;  S. Catani {\it et al.}, {\em ibid.} {\bf B264} (1986) 588 ; S. Catani, 
Z. Phys. {\bf C37} (1988) 357, and refs. therein.
\bibitem{gps} S. Jadach, B.F.L. Ward and Z. Was, Eur. Phys. J. {\bf C22}
(2001) 423.
\bibitem{ceex:2001}
S. Jadach, B.F.L. Ward and Z. Was, Phys. Rev. D{\bf 63} (2001) 113009.
\bibitem{elsewh} S. Jadach {\it et al.}, to appear.
\bibitem{bflw} B.F.L. Ward, Phys. Rev. D{\bf 36} (1987) 939.
\bibitem{sjw1990}Stanislaw Jadach and B.F.L. Ward, Comput. Phys. Commun.
{\bf 56} (1990) 351.
\bibitem{sjw1992} Stanislaw Jadach and B.F.L. Ward,
Phys. Lett. {\bf B274} (1992) 470.
\bibitem{BHLUMI-89}S. Jadach and B.F.L. Ward, Phys. Rev. D{\bf 40} (1989)
3582.
\bibitem{BHLUMI-92}S. Jadach {\it et al.}, Comput. Phys. Commun. {\bf 70}
 (1992) 305.
\bibitem{ward}B.F.L. Ward, Phys. Rev. D{\bf 42} (1990) 3249.
\bibitem{yfsww1} S. Jadach {\it et al.}, Phys. Rev. D{\bf 54} (1996) 5434.
\bibitem{BHLUMI-96} S. Jadach {\it et al.}, Comput. Phys. Commun. {\bf 102}
(1997) 229.
\bibitem{delaney}
D. DeLaney {\it et al.},Phys. Rev. D{\bf 52} (1995) 108, Phys. Lett. {\bf B342} (1995) 239.
\bibitem{tHvelt}G.~'t Hooft and M.~Veltman, Nucl. Phys. {\bf B44} (1972) 189 
and {\bf B50} (1972) 318, and references therein.
~\bibitem{nason} P. Nason {\it et al.},
Nucl. Phys. {\bf B303} (1988) 607; 
{\em ibid.}~{\bf B327}
 (1989) 49; {\em ibid.}~{\bf B335} (1990) 260.
\bibitem{been1} W. Beenakker {\it et al.}
Phys. Rev. D{\bf 40} (1989) 54; Nucl. Phys. {\bf B351}
 (1991) 507).
\bibitem{dglapsyn} B.F.L. Ward and Stanislaw Jadach,
Mod. Phys. Lett. {\bf A14} (1999) 491.
\bibitem{delaney2}D. DeLaney {\it et al.}, 
Phys. Lett. {\bf B292} (1992) 413.
\bibitem{mpa-ttbar}D. DeLaney {\it et al.}, Mod. Phys. Lett. {\bf A12}
 (1997) 2425.
\end{thebibliography}
\end{document}